# Contact angle entropy and macroscopic friction in non-cohesive two dimensional granular packings

Juan C. Petit P.
*Laboratorio de Física Estadística de Sistemas Desordenados, Centro de Física, Instituto Venezolano de Investigaciones Címtificas (IVIC), Apartado 21827, Caracas 1020 A, Venezuela*

Xavier García
*Schlumberger Geomechanics Center of Excellence, Gatwick, England.*

Iván Sánchez
*Research and Development Direction, Castillomax Oil and Gas S.A., Caracas, Venezuela.*

Ernesto Medina
*Yachay Tech, School of Physical Sciences & Nanotechnology, 100119-Urcuquí, Ecuador and Laboratorio de Física Estadística de Sistemas Desordenados, Centro de Física, Instituto Venezolano de Investigaciones Címtificas (IVIC), Apartado 21827, Caracas 1020 A, Venezuela*

We study the relationship between the granular contact angle distribution and local particle friction on the macroscopic friction and bulk modulus in non-cohesive disk packings. Molecular dynamics in two dimensions are used to simulate uniaxial loading-unloading cycles imposed on the granular packings. While macroscopic Mohr-friction depends on the granular pack geometric details, it reaches a stationary limit after a finite number of loading-unloading cycles that render well-defined values for bulk modulus, grain coordination, porosity, and friction. For random packings and for all polydispersities analyzed, we found that as inter-particle friction increases, the bulk modulus for the limit cycle decreases linearly, while the mean coordination number is reduced and the porosity increased, also as approximately linear functions. On the other hand, the macroscopic Mohr-friction increases in a monotonous trend with inter-particle friction. The latter result is compared to a theoretical model which assumes the existence of sliding planes corresponding to definite Mohr-friction values. The simulation results for macroscopic friction are well described by the theoretical model that incorporates the local neighbour angle distribution that can be quantified through the contact angle entropy. As local friction is increased, the limit entropy of the neighbour angle distribution is reduced, thus introducing the geometric component to granular friction. Surprisingly, once the limit cycle is reached, the Mohr-friction seem to be insensitive to polydispersity as has been recently reported.

## I. INTRODUCTION

The nonlinear mechanical behavior of granular matter is a subject of strong interest since it has many industrial applications, especially in structural and mechanical engineering, where concrete beams are designed and created to construct buildings and bridges more resistant to external load patterns [1–4]. This resistance is related to the shear strength and bulk modulus of the material. Both variables are also studied in soil mechanics using different mechanical assays such as triaxial and uniaxial tests [5–8]. The shear strength is the maximum shear stress that a material can sustain before failure. This maximum shear stress is related to the Mohr-Coulomb criterion where one can determine a macroscopic friction representing a resistance to shear stresses. The generalization of this criterion to non-cohesive granular packings with rigid and periodic boundaries in two dimensions [9–12] have suggested a relationship between particle friction, the contact angle distribution and the macroscopic friction of the granular sample. The relation between local and macro friction has also been explored in detail in three-dimensional packs [13–15], nevertheless the effect of the packing structure component on the macroscopic friction has not been widely explored.

The bulk modulus of a granular medium is a measure of the resistance to bulk deformation. There have been theoretical [16, 17], numerical [18–20] and experimental [21, 22] efforts to understand the macroscopic bulk modulus of a granular packing. However a general constitutive relation that predicts the global behaviour of granular system under external forces is still a challenge. Experimental [21, 22] and numerical [18] results report a scaling of the bulk modulus with pressure $P$ as $P^{1/2}$, in contrast to theoretical predictions [16, 17] and experiments [23], where a scaling of $P^{1/3}$ is found. This drawback was discussed in ref.[17], arguing that the assumption of affine deformations in mean field theories inevitably leads to the erroneous scaling. On the other hand, incorporating non-affine deformations makes the problem non-amenable to treatment with regular elasticity theory. Previous works have studied the effect of particle friction on the bulk modulus in 3D systems using isotropic compression [19, 20]. However, they focused on the elastic response of the system, when particle rearrangements are not present.

In this work, we will study numerically the effect of particle friction and polydispersity on the macroscopic friction and bulk modulus of granular packings. In sections II and III we describe our discrete particle dynamical simulations used to model uni-axial loading-unloading cycles applied to each granular packing. Such uni-axial test has been used to prepare isotropic packings [15, 44] and to reach stationary macroscopic properties [18]. Here, particle rearrangements and interpenetration are allowed to occur, and they result in changing granular pack properties from one loading-unloading cycle to the next. Our simulations allow for interpenetration values higher than those that appear in standard granular simulations, which are typically below of $\delta/R = 1\%$. Nevertheless, we show that there exists a stationary limit cycle above which the effective bulk modulus and Mohr-friction properties remain unchanged. It is this well-defined state of the granular pack that will be described, in order to clearly expose the geometric component as a determining factor to the macroscopic properties. In section IV the simulation results for the Mohr-friction will be compared with a theoretical model, that depends on the particle friction and the contact angle distribution within the pack. The contact angle distribution is quantified by the contact angle entropy that changes with local friction values and singles out the geometric component to Mohr-friction. As the local friction increases the contact angle entropy decreases and results in a larger Mohr-friction. This relation establishes a direct relation between granular order and macroscopic friction.

Finally, in section V, we study the effect of changing granular polydispersity on Mohr-friction and the bulk modulus. Previous works have studied the effect of polydispersity on the macroscopic friction in two and three-dimensional granular packings [15, 24], showing that the macroscopic friction was independent of this variable. Such independence was explained in ref.[24], as a consequence of the interplay between force chains and length scales within the packing. In these works triaxial and shear tests were used to impose a nominal number of stress states on the packing without reaching a limit hysteresis cycle. In this section, we address the relation between polydispersity and Mohr-friction in the loading-unloading stationary limit, finding that the latter is also surprisingly independent of polydispersity and is consistent with the theoretical model of the previous section. We end with a summary and the conclusions.

## II. CONTACT MODEL

When two grains of radii $R_i$, $R_j$, linear and angular positions $\mathbf{r}_i$, $\mathbf{r}_j$, $\theta_i$, $\theta_j$ come into contact, they interact with the following contact force [18, 25, 26]

$$\mathbf{F}_c = -\sqrt{\xi_n R_{ef}}(k_n\xi_n + \gamma_n\dot{\xi}_n)\hat{\mathbf{n}} + \mathbf{F}_s, \qquad (1)$$

where the first term is the normal contact force $\mathbf{F}_n$. The first term of $\mathbf{F}_n$ represents the elastic part of the model, with $\xi_n$ the relative normal displacement between the grain surfaces in contact, defined as $\xi_n = (R_j + R_i) - |\mathbf{r}_j - \mathbf{r}_i|$. $R_{eff} = R_iR_j/(R_i + R_j)$, is the effective radius between two grains, $k_n = 4\sqrt{E_{eff}}/3$ is the normal stiffness of the contact, $E_{eff} = E/2(1-\nu^2)$ is the effective Young's modulus at contact according to Hertz theory [27], while $\nu$ and $E$ is the Poisson's coefficient and Young's modulus of the grains. The second term of $\mathbf{F}_n$ represents the viscous part of the model, where $\gamma_n$ is the normal grain-grain damping coefficient, $\dot{\xi}_n$ corresponds to the overlap rate and $\hat{\mathbf{n}} = (\mathbf{r}_j - \mathbf{r}_i)/|\mathbf{r}_j - \mathbf{r}_i|$ is a unitary normal vector to the grain-grain contact. $\mathbf{F}_s$ represents the tangential contact force between the grains and it is modeled as in previous works [12, 18]. Such force is written in a compact form as

$$\mathbf{F}_s = -\min\left(\sqrt{\xi_n R_{ef}}\,(k_s\xi_s + \gamma_s\dot{\xi}_s), \mu_s|\mathbf{F}_n|\right)\hat{\mathbf{s}}, \qquad (2)$$

which depends on the minimum value between the viscoelastic force and the Coulomb condition for sliding grains. The first term represents the Mindlin model [28] with a viscoelastic contribution between two grains. This model is suitable for modeling tangential contact forces between grains and is widely recognized in literature [26, 29–31] although more rigorous models have been proposed [32] albeit more difficult to implement. $k_s = 8G/(2-\nu)$ is the tangential stiffness at the contact and $G$ is the shear modulus while $\gamma_s$ is the tangential damping coefficient. The second term in Eq.2 is the Coulomb friction force before sliding, where $\mu$ is the static microscopic friction between grains and $\xi_s$ is the relative tangential displacement between grains in contact computed as

$$\xi_s(t) = \int_0^t \dot{\xi}_s(t')dt', \qquad (3)$$

where $\dot{\xi}_s = \mathbf{v}_{ij}\cdot\hat{\mathbf{s}} + \omega_iR_i + \omega_jR_j$ is the magnitude of the relative tangential velocity between the center of the grains, $\omega_i$ and $R_i$ are the angular velocity and the radii of grain $i$. $\hat{\mathbf{s}} = \vec{\xi}_s/|\vec{\xi}_s|$ is a unit tangential vector to the surfaces in contact.

This model is used to simulate the contact between rotating and non-rotating disks. In the latter, the angular degree of freedom, $\theta$, is not considered allowing to sliding particles only. The comparison of the macroscopic friction between rotating and non-rotating disks are given in Figure 6. Additional simulations of rotating composite particles are also considered in order to explores the effect of angularity on their frustration of rotation. In Appendix B we will address this issue in detail.

TABLE I. Parameters used in simulation. Values correspond to quartz grains (see [33]), which are frequently found in sedimentary rocks.

| Prop. | Symbol | Value |
|---|---|---|
| Density | $\rho_g$ | 2.65 g/cm$^3$ |
| Normal stiffness | $k_n$ | 191.30 GPa |
| Tangential stiffness | $k_s$ | 183.32 GPa |
| Poisson ratio | $\nu$ | 0.08 |
| Damping coeff. | $\gamma_{n,s}$ | $2 \times 10^{-6}$g/(cm $\cdot$ s) |
| Micro friction | $\mu$ | [0.1-1] |
| Polydispersity | $\delta$ | [0-70]% |

## III. GRANULAR PACKING CONSTRUCTION AND SIMULATION PROCEDURE

A granular packing is composed of circular grains whose radii are in the range of $R \in [R_{\min}, R_{\max}]$. Such radii are chosen from a Gaussian distribution with average radius, $R_{av}$, and standard deviation $\sigma$, resulting in a polydispersity of the packing quantified by $\delta = \sigma/R_{av} = (R_{\max} - R_{\min})/(R_{\max} + R_{\min})$. Where $R_{\min}$ and $R_{\max}$ are chosen to correspond to $1\sigma$ around the mean of the Gaussian distribution.

The grain parameters used in simulations are given in Table I. The simulation box has the dimensions of $W = 100R_{\max}$ in width and $H = 200R_{\max}$ in height. The granular packing is constructed using a ballistic deposition algorithm, where grains fall into a position inside the box, one by one, where no frictional effects between grains are considered [34]. Once the packing is constructed, a vertical segment, $\Delta y = 10R_{\max}$, is taken at the bottom and at the top of the simulation box. Those grain positions satisfying $r_i \leq \Delta y$ belong to the bottom wall, while those satisfying $r_i \geq H - \Delta y$ belong to the top wall. These grain segments act as rigid blocks corresponding to pistons which are moved towards the center of the pack to compact the system. Figure 1 shows a typical packing and the corresponding pistons represented by the grey particles.

As was shown before [18], when one subjects the granular pack to a uni-axial loading, its configuration changes and the pack changes in length and properties, such as porosity, mean coordination number, bulk modulus etc. As it is inconvenient to study a non-stationary system under stress, we cycled the pack until no further changes occurred. This limit was typically reached for more than fifteen cycles, point at which one reaches a stationary hysteresis loop, i.e, an unchanging route in strain-stress space that closes on itself reproducibly. Each cycle consists of several stress states of compaction and closes on decompaction until it reaches the final state. Each stress state, during the loading process, is reached by moving the walls a distance $\delta y$ toward the center of the pack. After $N$ strain steps, the pack length changes in $\delta H = 2N\delta y$. The macroscopic strain $\epsilon_{yy}$ is calculated according to the relation

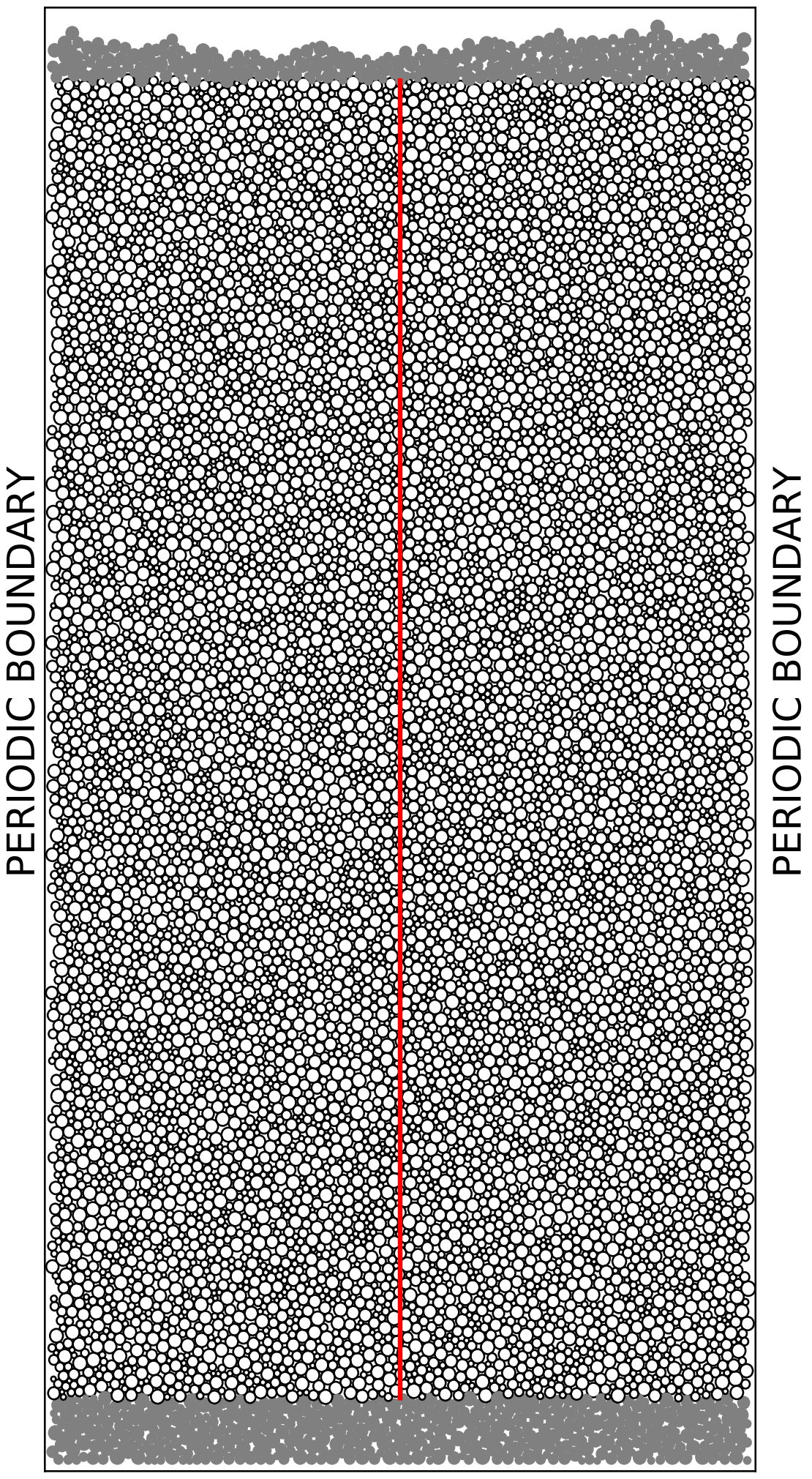

FIG. 1. Granular packing composed of 10900 grains constructed using a ballistic deposition algorithm. The polydispersity of the granular packing is $\delta = 50\%$. Grains colored in gray represent the pistons. Vertical black line represents an imaginary line used as guide to determine the horizontal stresses inside the packing.

$$\epsilon_{yy} = \frac{\delta H}{H}, \tag{4}$$

where $H$ is the reference length of the packing. During each stress state, part of the injected energy is stored as potential and kinetic energy of the grains, and the other part is dissipated through friction and damping at the grain contacts. Once the simulation reaches a particular strain state, it calculates the characteristic time to relax the kinetic energy of the grains to a predetermined value. Once this is verified, it is said that the system has reached a static configuration. At this point, the simulation continues with the next stress state. After a few steps, the packing reaches a maximum pre-selected vertical strain. Such value was considered here to be $\epsilon_{yy}^{\max} = 0.15$. Once the system reaches this value, the simulation starts the decompression process, where the pistons move away from the center of the granular packing, following the same procedure explained before. The decompression process ends when the stress on the pistons, $\sigma_{yy}$, falls below 5 MPa. The final configuration of each cycle will be the initial configuration of the following loading cycle. The $\epsilon_{yy}^{\max}$ value chosen leads to relatively large overlaps (see Appendix A) and long lasting contacts between disks. These cause larger mean coordination and lower porosity compared with those in standard simulations using hard-disk model [35].

In each strain state applied, the macroscopic stress $\sigma_{yy}$, transmitted along the vertical direction is calculated. In order to avoid wall effects on the macroscopic behaviour of the granular sample, we consider periodic boundary conditions in the direction perpendicular to compaction. An imaginary line is drawn along the vertical direction positioned in the center of the granular sample (see Fig.1) in order to determine $\sigma_{xx}$, which is calculated by a sum of the horizontal stresses that the grains on one side of the line exert on the grains on the other side. The whole procedure is described in ref.[18]. Once $\sigma_{yy}$ and $\sigma_{xx}$ are determined one can construct a Mohr circle through the following expression

$$\tau^2 + \left[\sigma_n - \left(\frac{\sigma_{yy} + \sigma_{xx}}{2}\right)\right]^2 = \left(\frac{\sigma_{yy} - \sigma_{xx}}{2}\right)^2, \quad (5)$$

that relates the shear stress, $\tau$, and normal stress, $\sigma_n$, for a particular plane inside the packing. The radius and center of each circle are calculated by $R = (\sigma_{yy} - \sigma_{xx})/2$ and $Ce = (\sigma_{yy} + \sigma_{xx})/2$ respectively. Eq.(5) corresponds to the case where $\sigma_{xx}$ and $\sigma_{yy}$ directions match with the principal axes of the packing. At the final cycle, we have a sequence of Mohr circles which are used to obtain the macroscopic friction by fitting a straight line as suggested by the Mohr-Coulomb criterion. Through this procedure, the normal and shear components applied instantaneously to every plane in the material will be described as a result of the uni-axial stress.

## IV. EFFECT OF INTER-PARTICLE FRICTION

### A. Bulk modulus

The bulk modulus of the granular pack is calculated following previous works [17, 18], where a variation of the vertical strain, $\Delta\epsilon_{yy}$, is imposed when monitoring the variation of vertical stress, $\Delta\sigma_{yy}$, and horizontal stress $\Delta\sigma_{xx}$. The bulk modulus here can be written as

$$K = \frac{\Delta\sigma_{yy} + 2\Delta\sigma_{xx}}{3\Delta\epsilon_{yy}}. \quad (6)$$

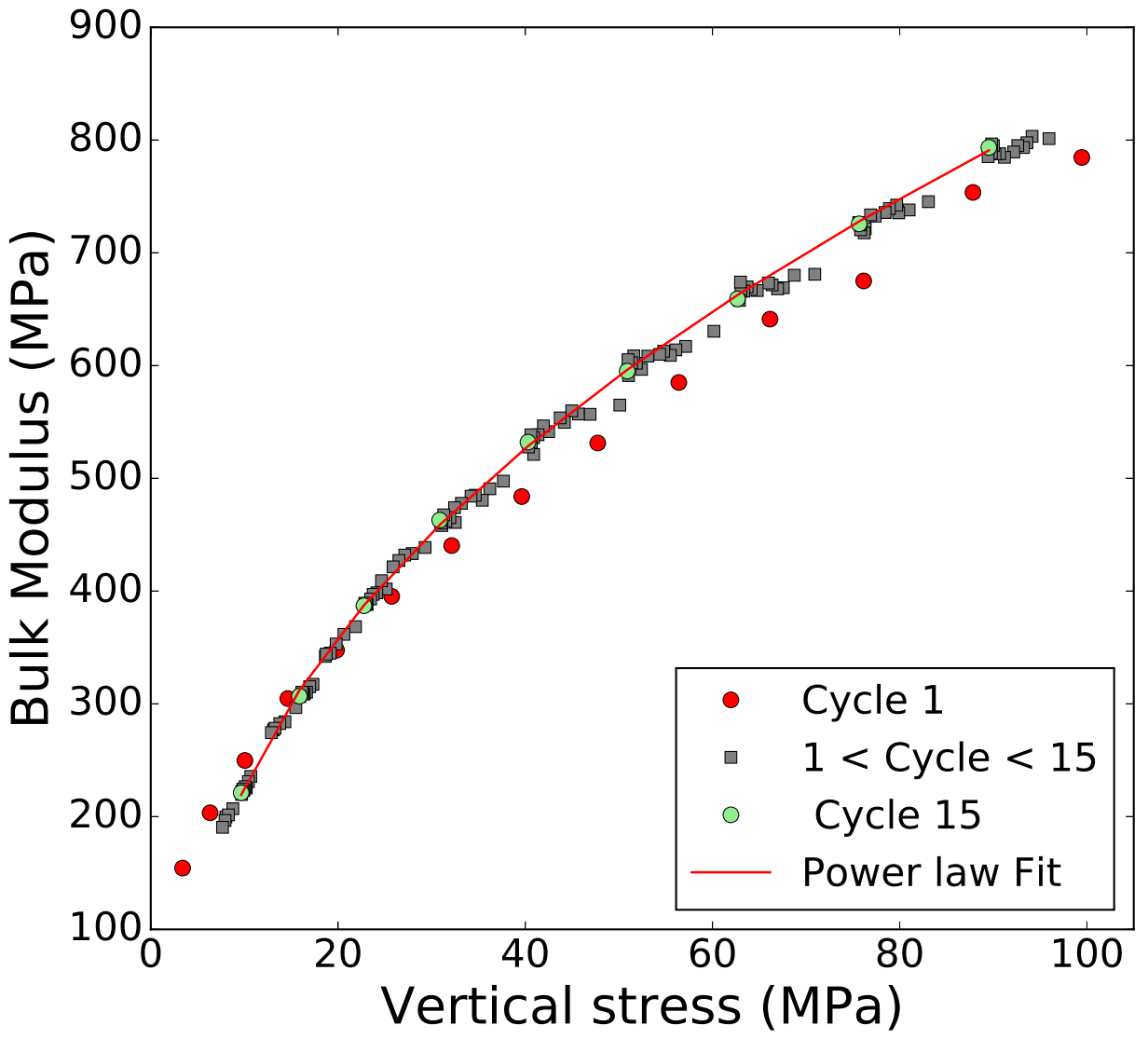

FIG. 2. Bulk modulus as a function of vertical stress for each cycle. The data is shown only for the loading process. A power law function is adjusted to cycle 15. Polydispersity and particle friction of the granular packing are $\delta = 50\%$ and $\mu = 0.3$ respectively.

Eq.(6) is strictly appropriate for macroscopically isotropic systems, while our system is anisotropic, especially due to how the contacts and the friction act redirecting the vertical pressure sideways. Nevertheless we use Eq.(6) as a measure of the bulk modulus, or how the system responds to changes in the volume of the granular pack in that particular direction. As we measure the stresses in the granular pack we can compute this quantity. We note that this relation and others related to macroscopic elastic quantities, such as shear and Young modulus, are used in soil mechanics assuming the structure of the soil as isotropic [36], that we know that this is not necessarily correct.

A typical curve of the bulk modulus as a function of vertical stress for each cycle during the loading process is shown in Figure 2. One can see that the granular pack hardens as the vertical stress increases. For low vertical stresses, the bulk modulus shows approximately the same values for each cycle. In this range, the packing behaves as an elastic system because particle rearrangements are less frequent. These results are consistent with recent works [19, 20], where the bulk modulus is addressed in such elastic range. For higher values of vertical stress, the bulk modulus increases with the number of cycles due to particle rearrangements for large deformations. Such rearrangements produce changes in the mean coordination number and porosity from the initial values: $Z_i = 3.8$ and $\psi_i = 17\%$, to the final values, $Z_f = 5.03$ and $\psi_f = 6.31\%$, for the final loading cycle. This cycle is taken as the limit cycle since variations of porosity and mean coordination number beyond the 15th cycle are less than $\pm 4.3\%$ and $\pm 0.2\%$ respectively.

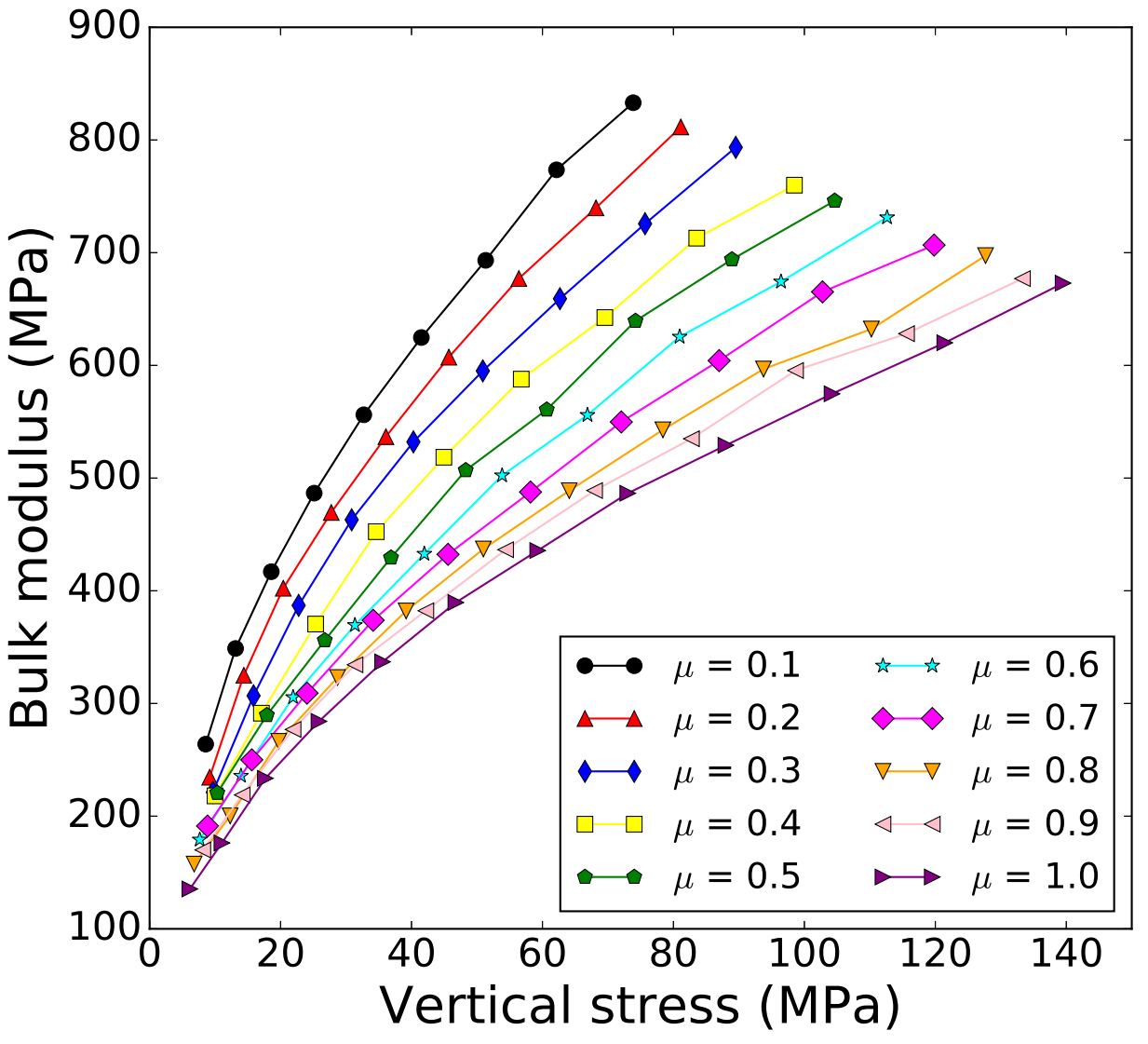

FIG. 3. Bulk modulus as a function of vertical stress with different particle frictions. The data is shown only for the loading process in the final cycle. The polydispersity of the granular packings is $\delta = 50\%$.

For the limit cycle, a good data fit can be made to a power law

$$K = K_0 + a\left(\frac{\sigma_{yy}}{\sigma_0} - 1\right)^{\alpha}, \tag{7}$$

where $K_0 \approx 200$ MPa is the elastic bulk modulus at minimum stress $\sigma_0 = 4.1$ MPa, $a = 0.29$ MPa and $\alpha = 0.47$. The value of the exponent obtained here is consistent with those found in experimental [21, 22] and numerical results [18], showing an exponent of 1/2 but above those obtained by mean field theories (MFT) [16, 17] and other experiments [23, 30, 37], where a 1/3 power law is found. This discrepancy between the theory, the experiments and simulations, is argued in ref.[17] to be due to the non-affine motion of grains during deformation, not taken into account in MFT. The latter approach taking into account non-affine deformations was attempted in refs.[38, 39], where they study the comparison of general aspects of non-affine behaviour evaluating correlations and spatial fluctuations of granular systems.

In order to study the effect of particle friction on the bulk modulus, we vary the particle friction in the range $\mu = [0.1 - 1]$. Ten granular packings were built with different particle frictions, with the same microstructure and a polydispersity of 50%. Then, fifteen uni-axial loading-unloading cycles were imposed for each packing. The bulk modulus was computed for each frictional value at the limit cycle using Eq.(6). Figure 3 shows the behaviour of the bulk modulus with vertical stress for different particle frictions when particle rotations are prevented. We obtain that, as particle friction increases the bulk modulus decreases for a specific applied vertical stress. These results are consistent with previous works in three dimensional granular packings (see refs.[19, 20]).

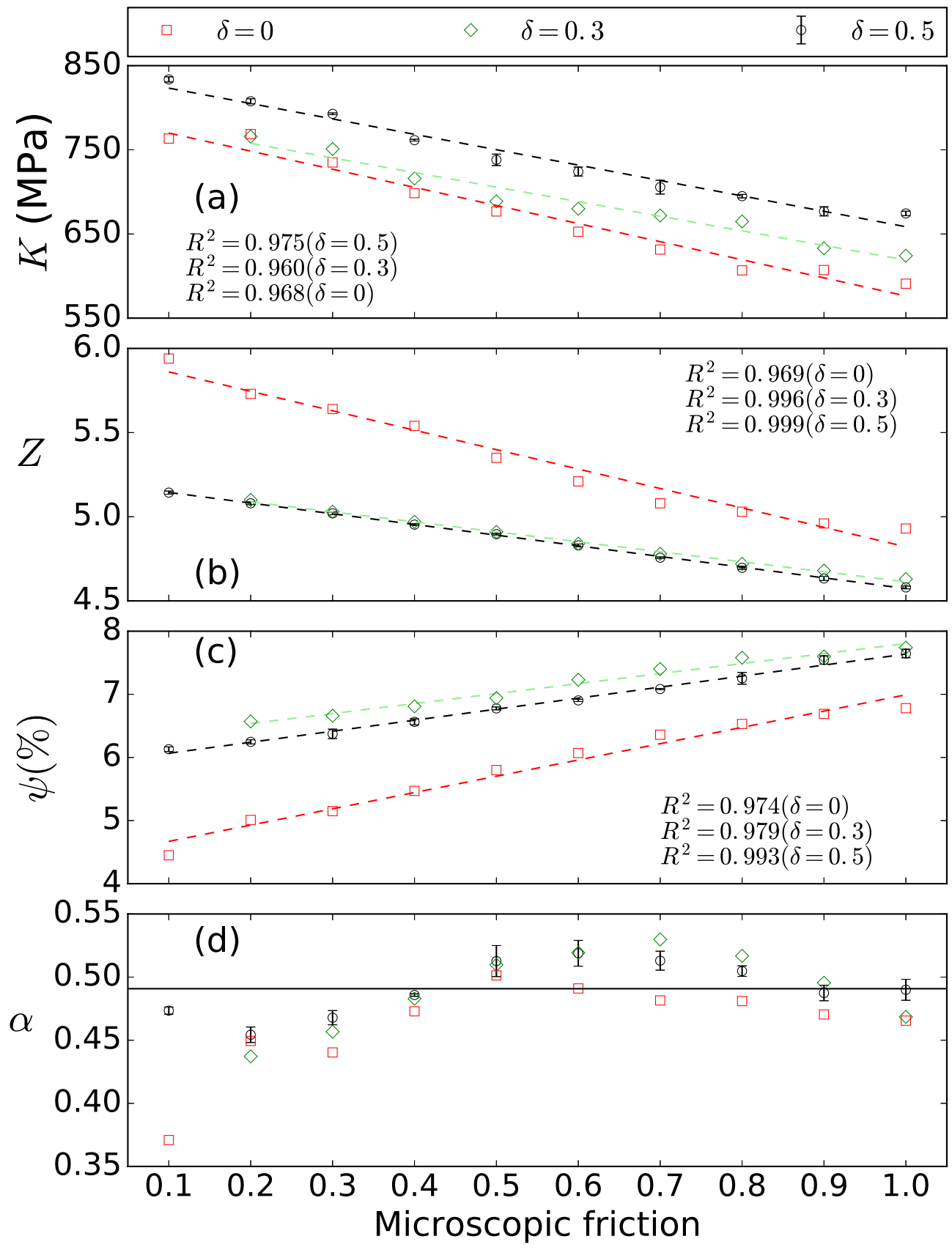

FIG. 4. (a) Bulk modulus, (b) mean coordination number, (c) porosity and (d) $\alpha$ exponent as a function of the microscopic friction. The data in (a) and (b) are shown for the final loading state of the final cycle, while the data in (c) are shown for the final cycle. Dashed lines represent linear fits to the data. Horizontal lines represent the mean value of $\bar{\alpha} = 0.49$. The data for the packing with $\delta = 0.5$ was averaged over 5 different samples respectively.

Figure 4a shows the reduction of the bulk modulus with particle friction for the final loading state of the limit cycle compared with the values obtained for less polydisperse and monodisperse packings. We can see that the three packings exhibit a non-monotonous linear trend with particle friction, where the highest polydisperse packing always shows larger bulk modulus values than the other packings.

The reasons behind the reduction of the bulk modulus with particle friction are related to change of the mean coordination number and porosity with this microscopic variable. Previous theoretical [16, 17] and numerical [19] results show a reduction of the bulk modulus when the mean coordination number decreases and porosity increases. Both variables are shown in Fig.4b and 4c as a function of particle friction, compared with those values obtained in polydisperse and monodisperse packs. We obtain that the mean coordination number decreases and porosity increases linearly with particle friction for the four packings. Such reduction is also obtained in a three-dimensional simulation [13]. At the grain level, increasing particle friction means less particle rearrangements, producing a reduction of the mean coordination number and as a consequence a less compact system of higher porosity. We note that, despite the high values of mean coordination number and low porosity exhibited by the monodisperse packing, it shows lower values for the bulk modulus as compared to polydisperse packings. This result suggests that the degree of polydispersity makes packings more rigid to bulk deformation than monodisperse packings of similar friction. We will address this point in section V.

We should note that the high and low values for the coordination number and porosity correspond to relatively high overlaps at the contact between grains. This can be clearly shown in Figure 14 of Appendix A, where we have assessed the distributions of contact interpenetration and the vast majority do not exceed 2%, although a few rare contacts can reach 8% for the highest polydispersity.

As an additional comment, the bulk modulus shown in Fig.3 were fitted to the power law given in Eq.(7) in order to explore the variation of the $\alpha$ exponent as a function of particle friction. We obtained that the $\alpha$ exponent fluctuates approximately 10% around its mean value of $\bar{\alpha} = 0.49$, see Fig.4d. Such variation meaning that the $\alpha$ exponent is insensitive to particle friction.

The results presented in this section confirm that the bulk modulus decreases with particle friction. This can be understood since packings with higher friction result in a lower coordination number and higher porosity. Friction interferes with the reshuffling of the granular pack into a more dense state.

### B. Mohr bulk friction

The macroscopic friction of the granular packing is determined extending the concept of Mohr's circles. This extension has been studied in two-dimensional granular systems in refs.[9, 11, 12], where the envelope of successive Mohr's circles, in the loading process, is found to be linear and is interpreted as an effective macroscopic friction for the granular sample. This straight line corresponds to the Mohr-Coulomb criterion written mathematically as

$$| \tau | = C + \mu_M \sigma, \tag{8}$$

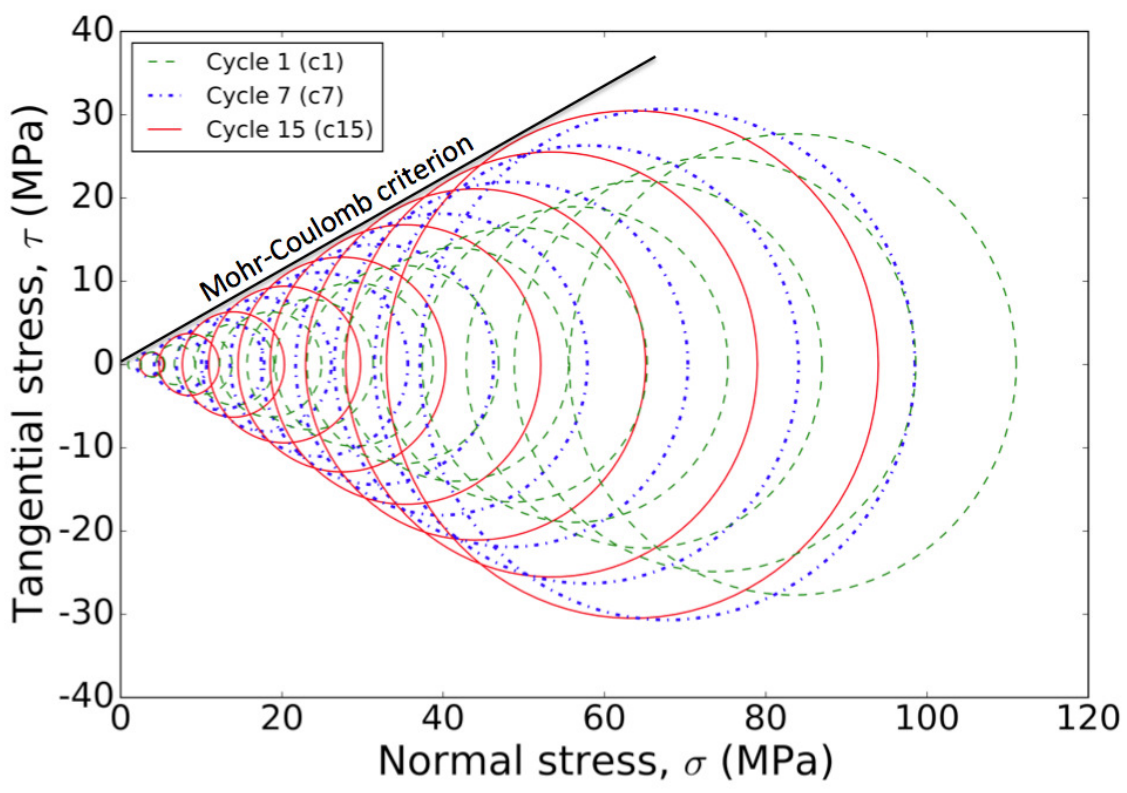

FIG. 5. Successive Mohr's circles in the loading process for the initial (c1), intermediate (c7) and final cycle (c15). The black line correspond with the fitting of the Mohr-Coulomb criterion. The polydispersity and particle friction of the granular packing are $\delta = 50\%$ and $\mu = 0.3$ respectively.

where $\sigma$ and $\tau$ are the maximum normal and shear stresses acting on a particular plane inside the material. $C$ is the shear stress due in general to cohesion of the granular pack (e.g. cementation between particles, not present here) and $\mu_M$ is the macroscopic friction of the packing. We focus on the second term, due to frictional stresses originated between particle friction and interlocking (geometrical effects). In addition, we study also how macroscopic friction changes with particle friction when particle rotations are considered or prevented. The latter case can be interpreted as if the packing was composed actually by grains with irregular shapes or angularity which contribute to frustrate their rotations.

Figure 5 shows successive Mohr's circles for increasing load imposed on a packing with $\mu = 0.3$. The innermost circles correspond with the first loading cycle, while the following two correspond to an intermediate and limit loading cycle. A macroscopic friction can be found for each loading succession of Mohr's circles given in Fig.5. The reference macroscopic friction for this packing is actually taken for the limit cycle since its value does not change more than $\pm 0.5\%$ for further cycles, so we have a stationary limit. For the limit cycle, we obtained $\mu_M \approx 0.5$, for both rotating and non-rotating disks, larger than the particle friction here at $\mu = 0.3$, consistent with previous works [9, 11, 12]. This result is expected, since the dilatancy of the material depends on the structure of the packing and opposes the sliding of fault zones [40] that are contemplated in the Mohr-Coulomb criterion.

The ten packings with different particle friction and the polydispersity of $\delta = 50\%$ were used to determine their macroscopic friction as a function of $\mu$. Figure 6 shows this relation considering the case of both sliding and particle rotation, and only particle sliding. We can see that when particles only slide the macroscopic friction increases approximately linear with $\mu$, evidencing values always larger than this microscopic variable. However, when particles are able to slide and rotate we observe two different behaviours for the macroscopic friction. One regime in which $\mu_M > \mu$ and another where $\mu_M < \mu$. We can interpret such behaviours as a competition between particle sliding and particle rotations. For the smaller $\mu$ values, points follow the curve considering only sliding, suggesting that sliding is dominant over rotations. Particle rotations start to contribute as $\mu$ increases and become dominant for larger values of local friction. This leads to the saturation range which has been also obtained in bidimensional [11, 43] and threedimensional simulations [15, 44]. This clearly demonstrate that 2D simulations are able to reproduce important macroscopic properties, such as macroscopic friction, exhibited in 3D packings.

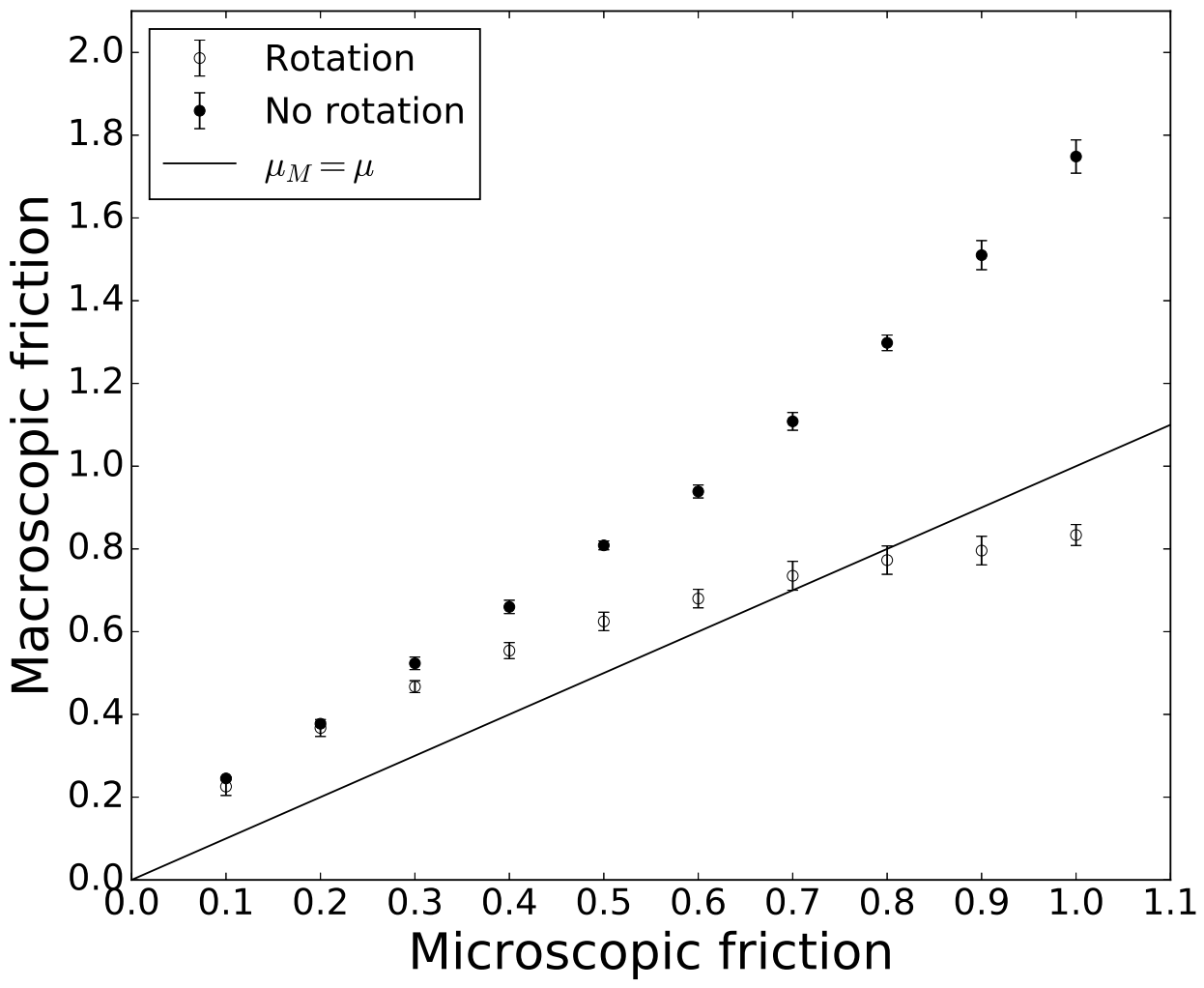

FIG. 6. Macroscopic friction as a funtion of particle friction. Hollow symbols represent the data for cycle 1 with slidings and particle rotations. Filled symbol are the data for cycle 15 with only particle slidings. The degree of polydispersity for the packings was set to $\delta = 50\%$.

The simulation results obtained for the case of particle slidings are compared with a simple bidimensional model which relates friction and contact orientation of grains with the macroscopic friction of the packing. This model was developed in ref.[9] using Rankine analysis under the assumption that grains can only slide but no rotate, and tested by numerical simulations presented in ref.[12].

Although particle rotations in granular simulations are crucial since they control e.g. the development of shear bands in granular packings [45], frustration of particle rotations seems to be the norm when more realistic angular particles are considered and higher overlaps occur [46, 47]. Our simulations, that allow for some overlap beyond 1% can bind disks together obstructing rotations. This situation is more in tune with a granular solid than a granular material and possibly a transition between the two [48]. This statement allows us to justify the non-rotating disk model used in our simulations, which can be viewed as a limiting case for angular grains frustrating all rotations. To show this, we undertook an independent validation of our rotationally frustrated limit by modelling more realistic angular particle systems as particles made up of clumps of disks with varying "angularity". The results, described in detail in the Appendix B, show an increasing dominance of sliding for a larger range of local friction values as such angularity increases. These results are consistent with previous results [49, 50], where mean friction mobilization increases as angularity increases. Based on this fact, our theoretical model can be understood as one addressing the contribution of rotation frustration by the angularity of the grains that increase the shear strength of the packing. For the rest of the paper, we will ignore rotations in our granular pack treating only this limit.

The macroscopic friction in this model can be calculated using the following expression derived in ref.[12]

$$\langle \mu_M \rangle = \int_{\phi_{\min}}^{\phi_{\max}} P(\phi) \tan\left(\frac{2}{3}\left[\phi_m - \phi + \pi/4\right]\right) d\phi, \quad (9)$$

where $P(\phi)$ is the contact angle distribution in the packing, $\phi_m$ is the friction angle between grains so that $\tan \phi_m = \mu$. $\phi_{\max}$ is the maximum angle of contact allowed given by $\phi_m + \pi/4$. This limit is found by imposing that $\tan \phi_M \geq 0$. Nevertheless, the lower limit $\phi_{\min}$ is a free parameter, depending on the structure of the sliding plane, and is further restricted. For example, for a perfectly ordered straight line of spheres of the same size, the lower limit is $-\pi/6$. Roughness of such a plane or line (for a two-dimensional pack), will yield larger average angles [9].

Figure 7 shows the comparison between a normalized contact angle distribution of the packings before compaction and some of them after the final loading cycle for different particle friction. Before compaction, the initial distribution exhibits an anisotropic structure (top panel) with privileged orientations between grains. Less frequent contact angles are about 0°, which corresponds to one grain on top of the other, and $\pm\pi/2$ which is for grains to the left or right of each other. Interestingly, after the limit number of cycles, the distributions change depending on the local friction value. Such change also depends on the number of cycles performed. For low inter-particle friction the distribution is quite isotropic. However, for intermediate and higher inter-particle friction, the distributions show highly privileged angles around 0°, ±30° and ±90°.

To study the concentration of these distributions as a function of the number of stress states imposed, we use the definition of Shannon entropy, written as

$$S = -\sum_{k=1}^{N} P(\phi_k) \ln P(\phi_k), \quad (10)$$

TABLE II. Values of $\phi_{\min}$ used to integrate equation (9) for the initial (C1) and final cycle (C15). (Neg) was written for short and it means that all values of $\phi_{\min}$ are negatives.

| $\mu$ | 0.1 | 0.2 | 0.3 | 0.4 | 0.5 | 0.6 | 0.7 | 0.8 | 0.9 | 1.0 |
|---|---|---|---|---|---|---|---|---|---|---|
| $\phi_{\min}(C1)$(Neg) | 22.9° | 25.7° | 31.5° | 35.5° | 39.5° | 39.5° | 39.5° | 39.5° | 39.5° | 39.5° |
| $\phi_{\min}(C15)$(Neg) | 25.7° | 37.2° | 45.8° | 48.7° | 48.7° | 48.7° | 46.9° | 46.4° | 45.8° | 42.9° |

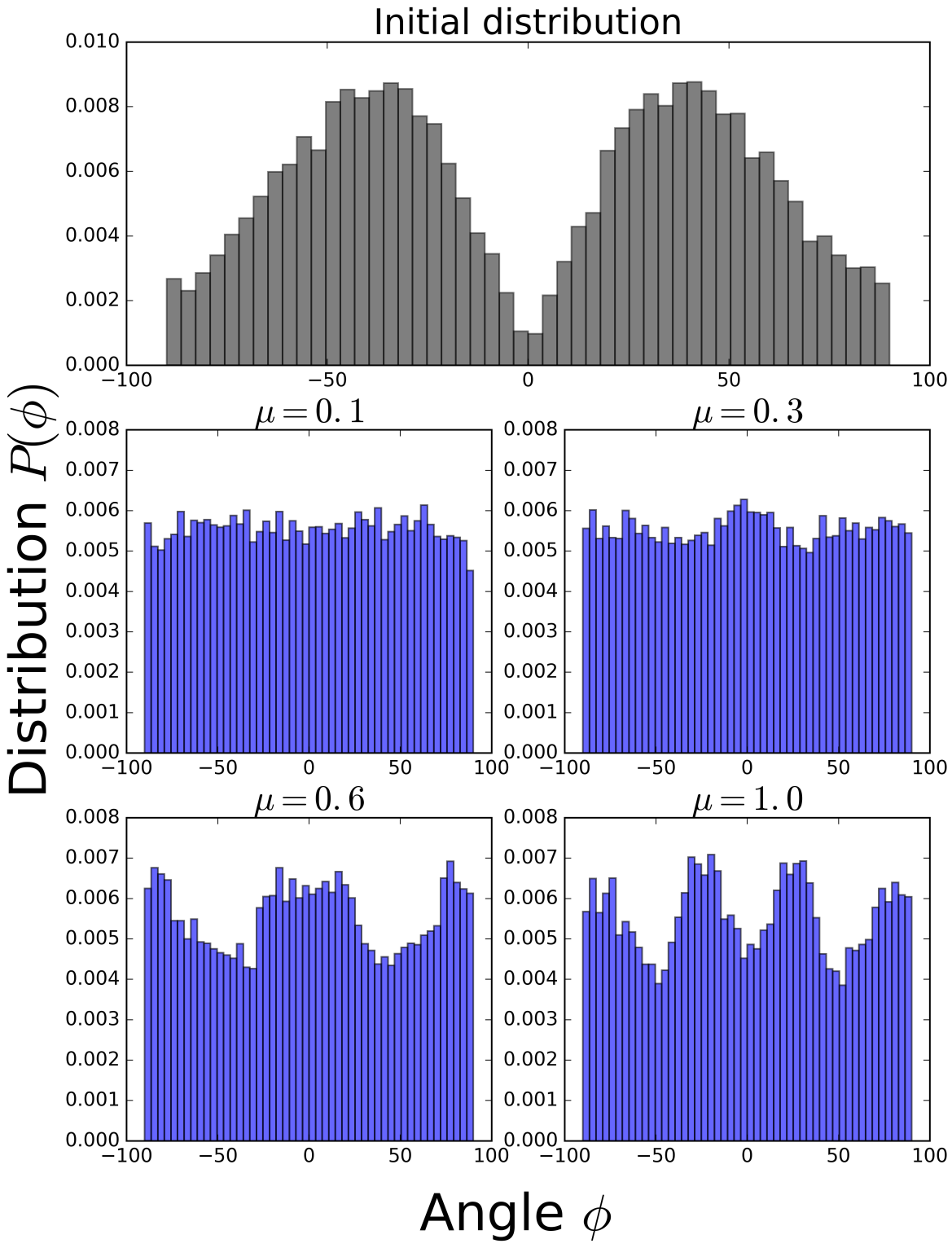

FIG. 7. Comparison of the initial contact angle distribution of the packings before compaction and four distribution after final loading cycle for different particle friction. The polydispersity of each granular packing is $\delta = 50\%$.

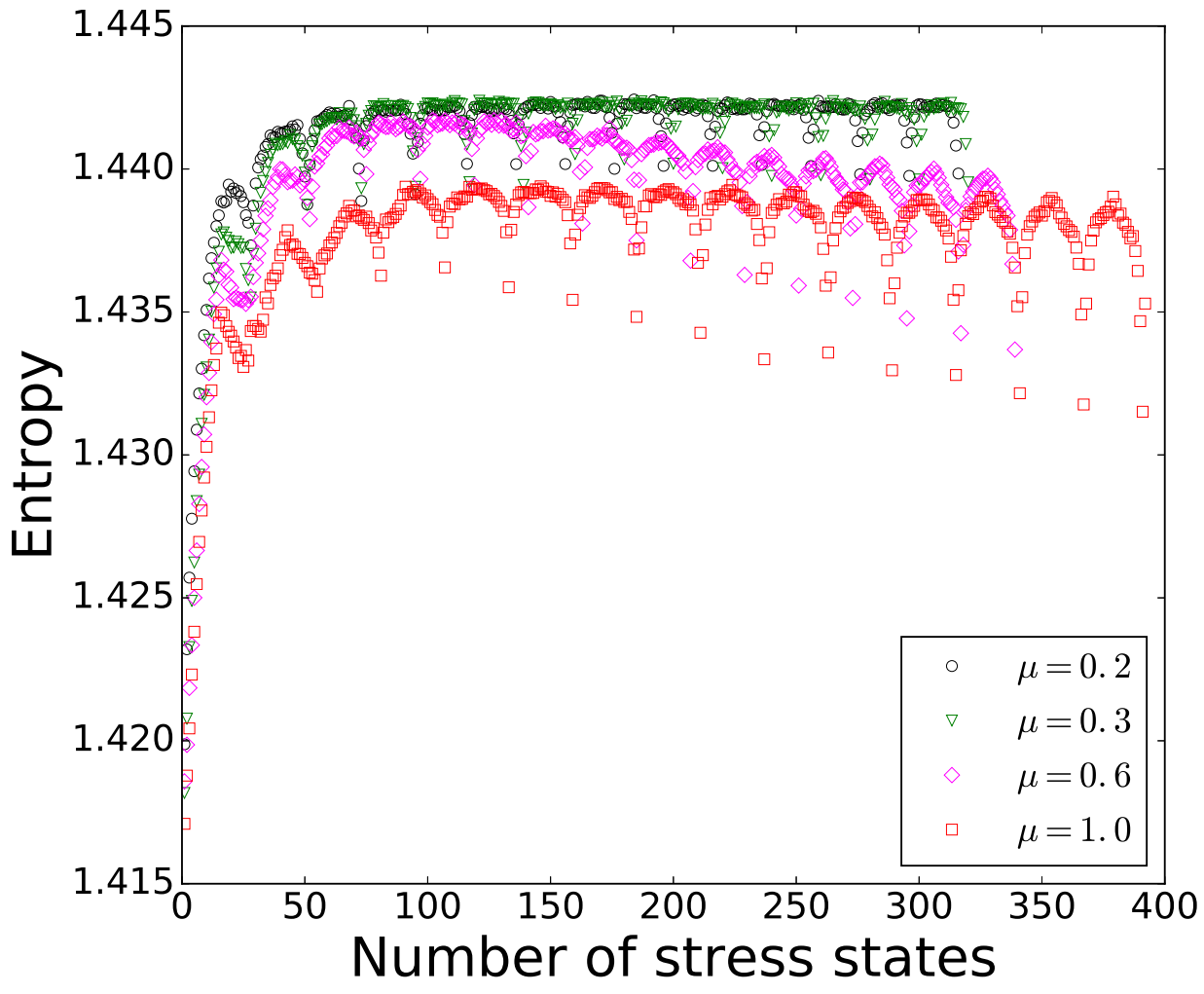

FIG. 8. Shannon entropy as a function of the number of stress states applied to four packings, with a degree of polydispersity of $\delta = 50\%$ and different particle frictions.

where the sum is over the number of $\phi_k$ angles exhibited by the distribution. Such an expression has been used in the literature as a measure of the structural disorder of the contact angles in the packing [51, 52]. Figure 8 shows the initial value of the entropy, $S \approx 1.415$, which corresponds to the first loading state applied on the packings. This value corresponds to the initial distribution given in Fig.7, representing a quite ordered structure. The latter value of entropy is higher than the entropy calculated for a triangular packing, $S_{\text{tri}} \approx 1.386$, representing the most ordered packing. Successive loading states make the packings evolve from an ordered to a disordered state, increasing their entropy until the final loading state of the first cycle is reached. After this point, the unloading process begins and the entropy decreases slightly, suggesting a small reduction of disorder inside the packing. Successive loading-unloading cycles increase the entropy, until an approximated saturated value is reached. Such value depends on particle friction. For example, for $\mu \leq 0.3$ the packings exhibit a maximum value of entropy in the limit cycle. This value correspond to a uniform distribution, as we can see in Fig.7, defining a disordered packing. As particle friction increases, $\mu > 0.3$, displacement of particles is frustrated by friction and the packing reaches a lower maximum entropy. In this case, there remain privileged contact angles, defining an ordered packing.

An interesting feature is that while less frictional packings reach a stable value for the maximum entropy, the more frictional packings first reach a peak for the entropy, which then gets reduced for further cycles. We have no simple explanation for this effect, but it seems that the contact angle entropy is a sensitive way to probe granular pack changes.

The evolution of the contact angle distribution with loading cycles has not been contemplated before as a determining factor of Mohr-friction. In fact, previous attempts to fit the theory to simulation, in ref.[12], assumed as a good approximation the initial uncycled contact distribution for all friction values, while in ref.[9], two ad hoc distributions where surmised. These assumptions are clearly incorrect in view of the previous discussion on the results of Fig.7, so that, it is necessary to use the distribution for each local friction to evaluate Eq. (9). Figure 9 shows a comparison between the theoretical relation involving Eq.(9) and the Mohr-friction. We find that the model fits very well the simulation results, representing a linear trend. This result was achieved by using a particular value of $\phi_{\min}$ for each distribution with a specific particle friction. Such values, listed in table II, represent the best for adjusting the theoretical model to the simulation results. Additionally, the theoretical model was also adjusted to the simulations results for the first cycle and we obtained a good fit once the appropriate distributions were considered. The values of $\phi_{\min}$ obtained for this cycle are those given in table II. We highlight that the values of $\phi_{\min}$ for cycle 15 are larger than cycle 1, increasing slightly with particle friction. As particle friction increases both $\phi_{\max}$ as $\phi_{\min}$ increase, this means that the range of contact angles to evaluate Eq.(9) is wider, contributing to a higher macroscopic friction, (see Fig.9). We can see that the macroscopic friction increases within the range of inter-particle frictions considered, being always larger than $\mu_M = \mu$. A linear fit is also showed as a guide to the eye.

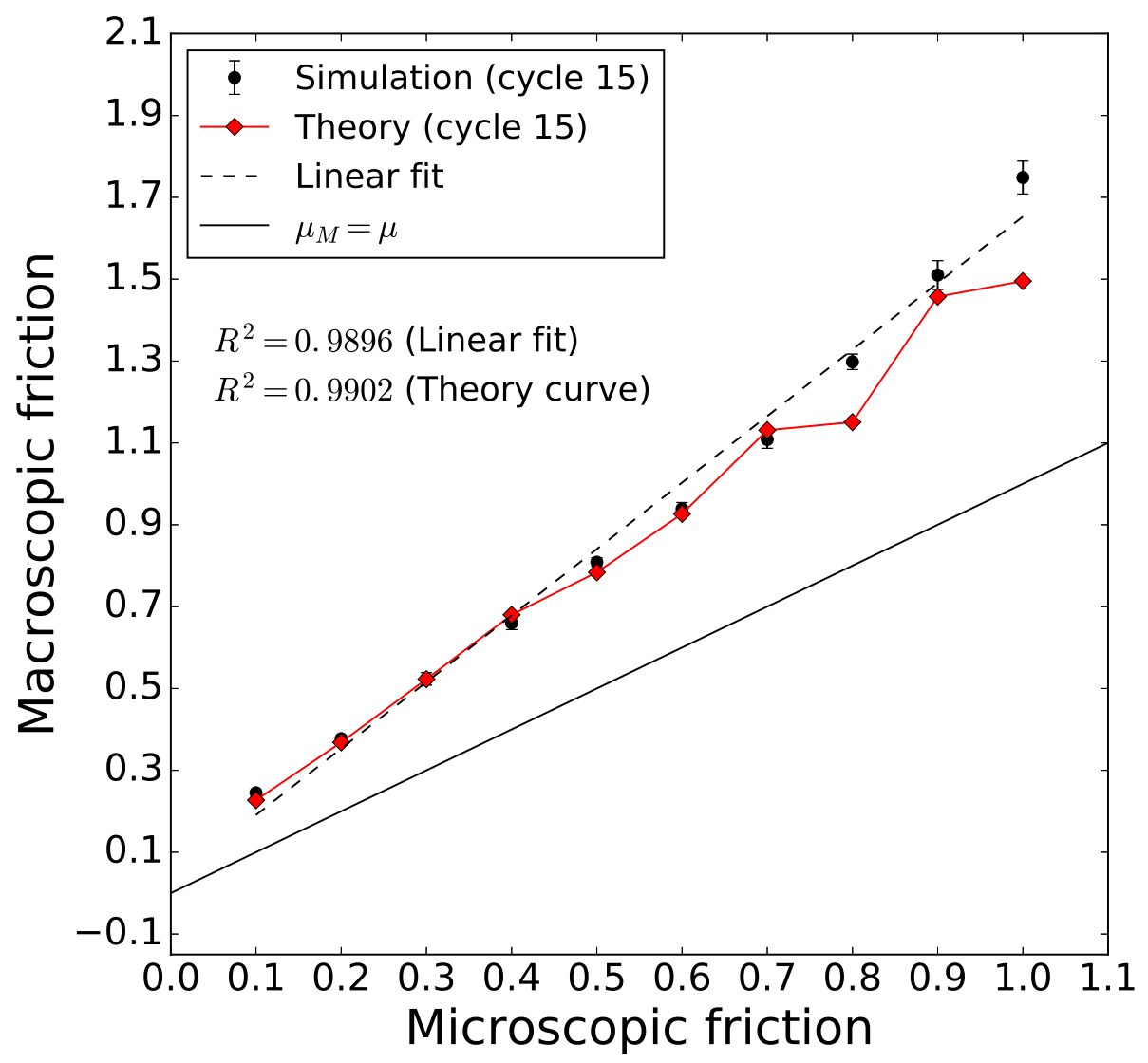

FIG. 9. Macroscopic friction as a function of particle friction for the final loading cycle. The polydispersity of the granular packings is $\delta = 50\%$.

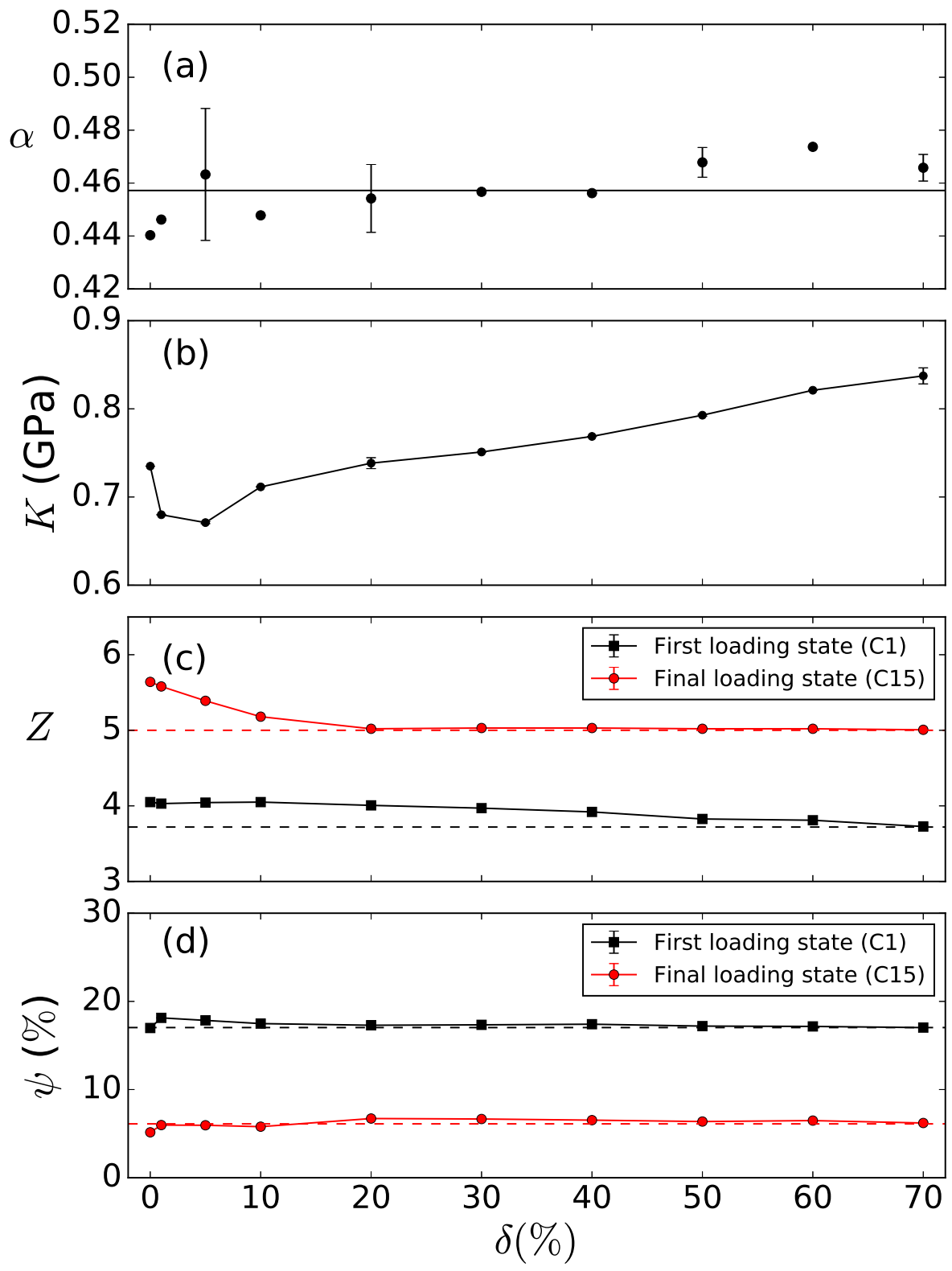

FIG. 10. (a) $\alpha$ exponent, (b) bulk modulus, (c) mean coordination number and (d) porosity as a function of polydispersity. Square symbols (■) correspond to the data for the first loading state of the first cycle (C1), while circular symbols (•) correspond to those for the final loading state for the final cycle (C15). Horizontal dashed lines represent guide lines. Horizontal solid line represents the mean value of $\bar{\alpha} = 0.458$. Particle friction was set to $\mu = 0.3$.

## V. EFFECT OF POLYDISPERSITY

### A. Bulk modulus

The previous studies were done at a fixed polydispersity, so it is necessary to gauge the robustness of those results against changes in polydispersity. Ten packings with different polydispersity in the range $\delta = [0 - 70]\%$ are created, keeping the same particle friction $\mu = 0.3$. After applying a sufficient number of uni-axial loading-unloading cycles in order to reach the limit stationary pack, the bulk modulus was calculated using Eq.(6). The bulk modulus for each packing increases with vertical stress following a power law of the form $K \sim \sigma_{yy}^{\alpha}$. The exponent $\alpha$ as a function of polydispersity is shown in Fig.10a, where its value changes less than 4% with respect to its mean value $\bar{\alpha} = 0.458$, so that $\alpha$ is independent of polydispersity within the error bars. Only the data of $\delta \in [5, 20, 50, 70]\%$ were averaged over three different samples.

Figure 10b shows the values of the bulk modulus for the final loading state of the final cycle as a function of polydispersity. The bulk modulus increases linearly with polydispersity in the range of $\delta \in [5, 70]\%$. This result is in contrast to those obtained in compressional three-dimensional granular packings [41], where the elas-tic modulus decreases as the degree of polydispersity increases. Figures 10c and 10d show the values of the mean coordination number and porosity, compared with those of the loading stage of the first cycle as a function of polydispersity. The figures show that the mean coordination number and porosity decrease slightly with polydispersity. After the final loading state of the final cycle, the mean coordination number has increased with respect to the values of the first cycle but still decreases with polydispersity very weakly. This result is consistent with previous work in three-dimensional systems [41]. On the other hand, the porosity decreases respect to the values of the first cycle, but it varies little with polydispersity, showing an approximate saturation value for each packing.

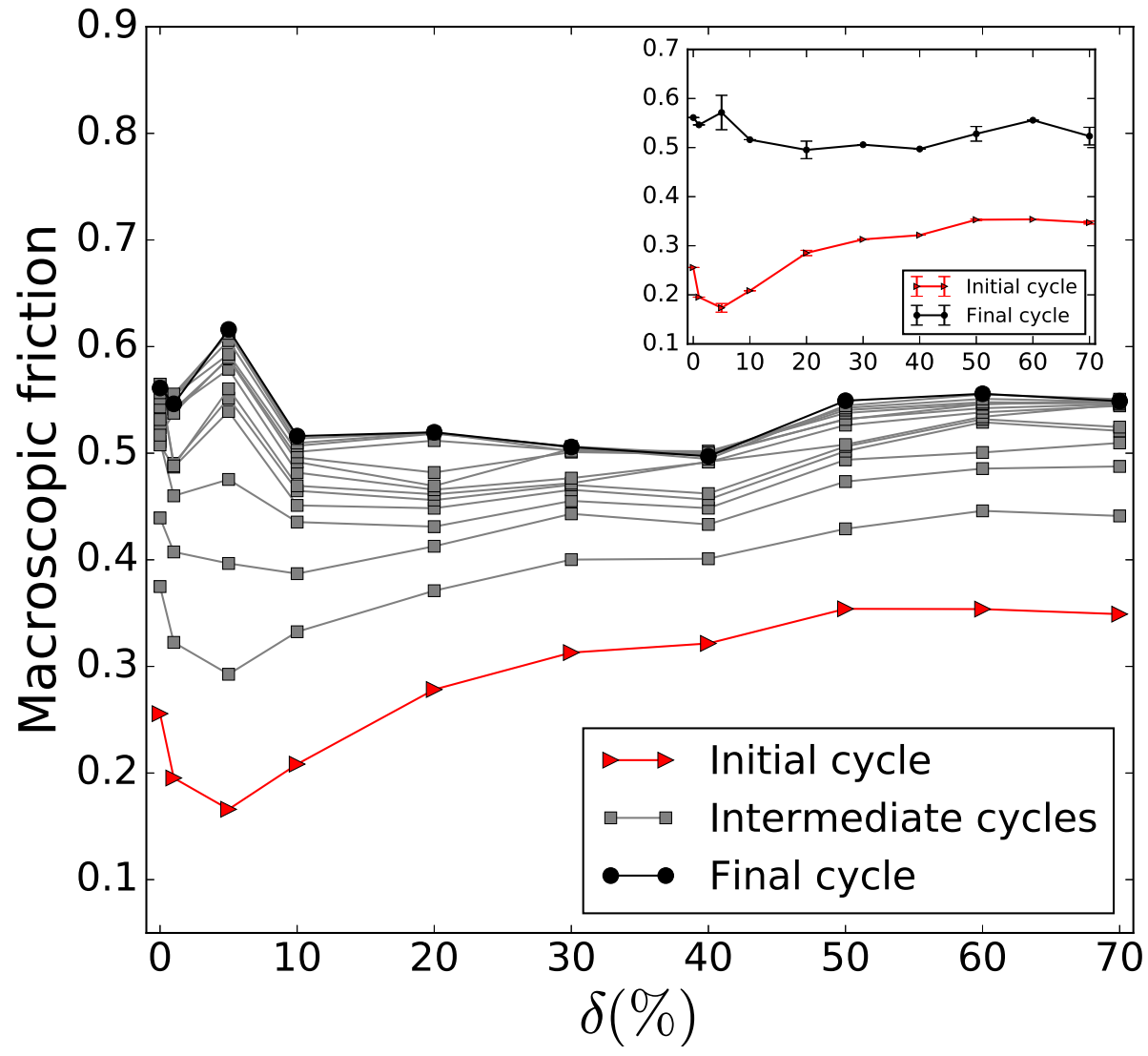

FIG. 11. Macroscopic friction as a function of polydispersity for each cycle imposed. The particle friction of the packings is $\mu = 0.3$. In the inset we show the curves of macroscopic friction with polydispersity for the initial and final cycle with four error bar corresponding to $\delta \in [5, 20, 50, 70]$. The data were averaged over three different samples.

In Fig.10b we can see that the packing with the highest bulk modulus is that with the highest polydispersity, showing lower mean coordination number and the same porosity as monodisperse or slightly polydisperse packings. This is an unexpected result since previous works have shown that the mean coordination number of packings decreases with the degree of polydispersity [31, 41], being responsible for the reduction of the bulk modulus [19]. We would have expected that packings with higher polydispersity would exhibit a lower bulk modulus than monodisperse or slightly polydisperse packings however this is not the case. Preliminar results show that while the degree of polydispersity increases, the number of large grains increases, a fraction of them support strong forces which form a rigid structure inside the packing difficult to overcome during deformation. This argument is similar to previous results in highly polydisperse packings composed of disks [24] and pentagonal grains [42], where strong forces propagate through more larger particles as the size of polydispersity increases. We think that such structures are responsible primarily for the increment of the bulk modulus with polydispersity. These findings will be part of a future work.

### B. Mohr-friction with polydispersity

Figure 11 shows the relation of the macroscopic friction with polydispersity for each cycle imposed, setting $\mu = 0.3$ for all polydisperse packings. We see that the macroscopic friction for different polydispersities increases with cycle, obtaining a saturation value in the final cycle. For the first three cycles, we obtain that increasing polydispersity increases the Mohr-friction. After a number of cycles, nevertheless, the macroscopic friction is essentially constant, showing only statistical variations around the mean value of $\bar{\mu}_M = 0.54$. These results suggest that the macroscopic friction is independent of polydispersity beyond the limit cycle and it achieves this behavior by the reorganization of the pack. A previous work [15], studied the macroscopic friction with polydispersity in three-dimensional packings. They found that the macroscopic friction increases from 0.41 for lower polydispersity to 0.44 for higher ones implying only a weak dependence. Other works on highly polydisperse packings studying force chains and macroscopic friction of disks [24] and pentagonal grains [42] found that the macroscopic friction was independent of the size of polydispersity but unexpectedly, it declines with increasing the degree of shape irregularity of pentagonal grains.

The reorganization of the grains inside the packings was explored using Eq.(10). Figure 12 shows how the initial entropy increases as a function of polydispersity, showing that packings with higher polydispersity have more disordered structures. On the other hand, the monodisperse packing exhibits the lower entropy (see inset in Fig.12), displaying a strongly ordered structure. In the final loading state of the last cycle, packings of $\delta \geq 30\%$ show almost the same value of entropy, suggesting that an equivalent configurational structure was reached after a given cycle, i.e., similar contact angle distribution were obtained. Nevertheless, for packings of $\delta < 30\%$, the entropy reaches a maximum value for a particular stress state, and then it reduces to an approximate stationary value. Even though a different contact angle distribution was obtained for slightly polydisperse or monodisperse packings, a similar value for the Mohr-friction were obtained (see Fig.11). This is an interesting kind of universal behaviour (independence to a degree of polydispersity) that should explored further.

On the other hand, changing the value of microscopic friction from $\mu = 0.3$ to $\mu = 0.7$, we also obtained dif-ferent results on the curve of entropy as a function of the number of stress states for different polydispersities, (see Figure 13). Here we also see the behaviour observed in the previous section (see Fig.8) where the entropy changes qualitatively below $\mu = 0.3$. We obtained that for all polydispersities considered the entropy reached a maximum value to then relax to a lower value. When particle friction increases the contact between grains are preserved since particle rearrangement are less frequent. From this point of view, polydisperse packings are effectively more ordered than those with $\mu = 0.3$, obtaining smaller entropy values. An opposite result is obtained for monodisperse packing, where a more disordered state is reached when $\mu = 0.7$ than that with $\mu = 0.3$.

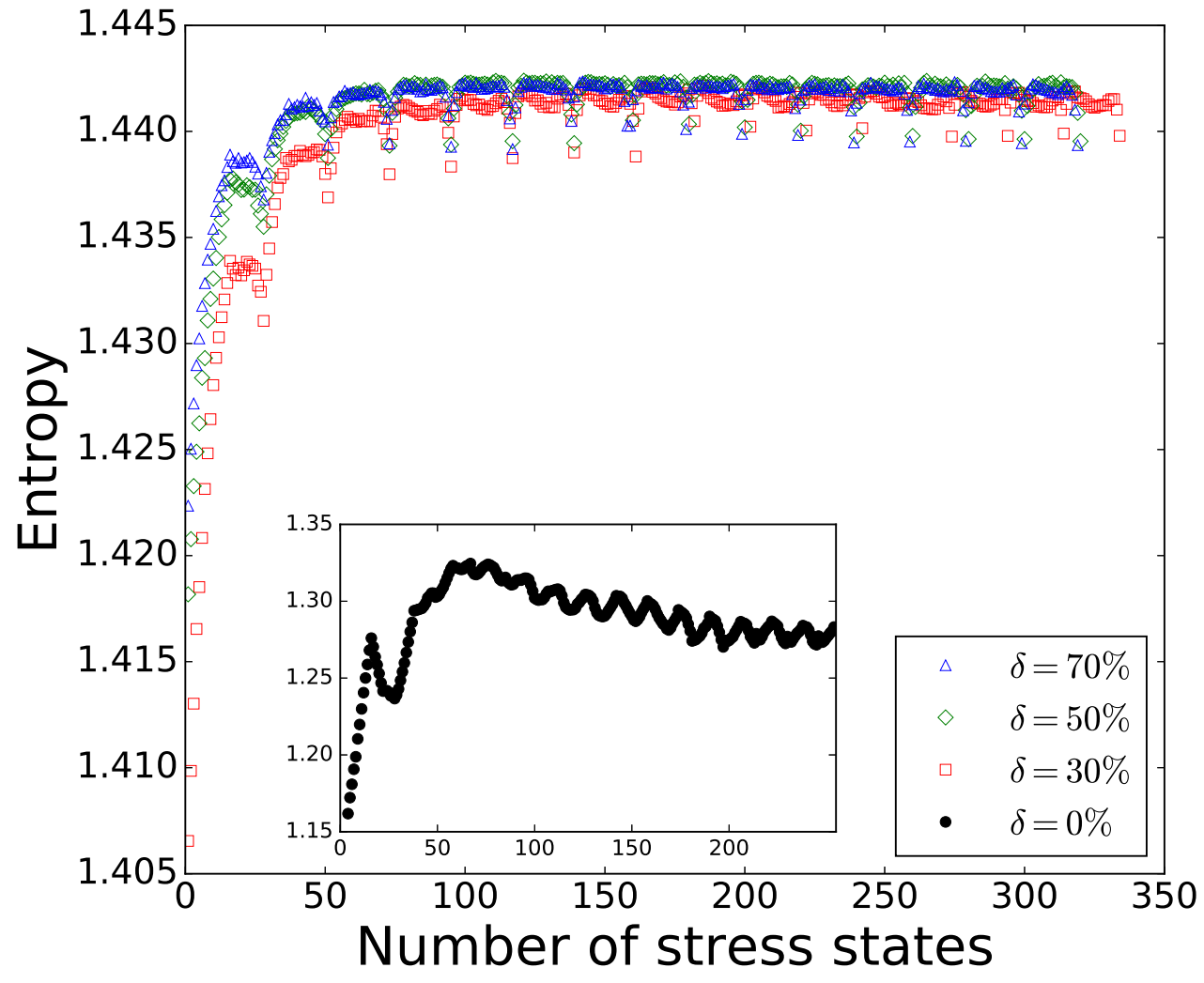

FIG. 12. Shannon entropy as a function of the number of stress states for four packings with different degree of polydispersity. The Inset shows the Shannon entropy values for monodisperse packing. Particle friction for each packing is $\mu = 0.3$.

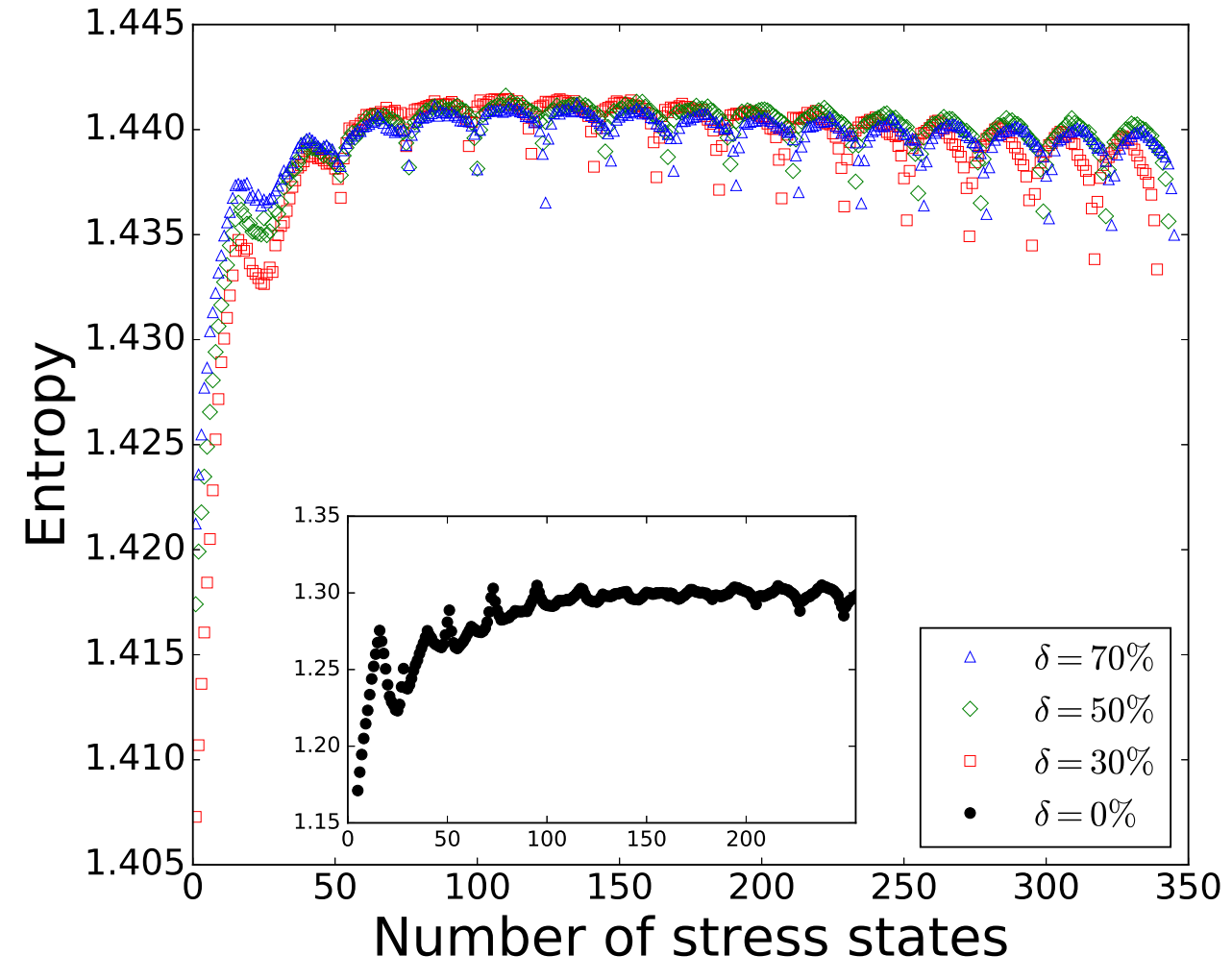

FIG. 13. Shannon entropy as a function of the number of stress states for four packings with different degree of polydispersity. The Inset shows the Shannon entropy values for monodisperse packing. Particle friction for each packing is $\mu = 0.7$.

## VI. SUMMARY AND CONCLUSIONS

We have studied the relationship between the granular contact angle distribution and local particle friction on the macroscopic friction and bulk modulus in two dimensional non-cohesive granular packings. The system studied is a granular pack subjected to uni-axial loading-unloading cycles. We found that the system has reached a limit cycle where its properties remain stationary under uni-axial stress. For random packings and for all polydispersities analysed, we found that as inter-particle friction increases, the bulk modulus for the limit cycle decreases linearly, while the mean coordination number is reduced and the porosity increased, also as approximately linear functions. On the other hand, the macroscopic Mohr-friction increases in a monotonous trend with inter-particle friction.

Quantifying the geometrical structure of the cycled granular pack through the contact angle distribution, we find that it depends critically on the local friction values displaying a multiply peaked distribution for the larger friction values. This is well evidenced through the values of the contact angle entropy, showing how the pack is organized as it is compressed and cycled. The Mohr-friction trend is compared to a theoretical model which assumes the existence of sliding planes corresponding to definite Mohr-friction values. The simulation results for macroscopic friction are well described by the theoretical model only when the details of the particular neighbour angle distribution is contemplated. As local friction is increased, the limit entropy of the neighbour angle distribution is reduced, thus demonstrating the geometric component to granular friction. Surprisingly, once the limit cycle is reached, the Mohr-friction seem to be insensitive to polydispersity as has been recently reported. The latter behaviour is also seen for the contact angle distribution and entropy which is practically unchanged as a function of polydispersity. Thus contact angle entropy seems to be a useful tool to assess the geometrical contributions to granular pack Mohr-friction.

## ACKNOWLEDGMENT

The authors gratefully acknowledge illuminating discussions with Vanessa Urdaneta. EM acknowledges support from IVIC through the “Granular Dynamics Project” 271.

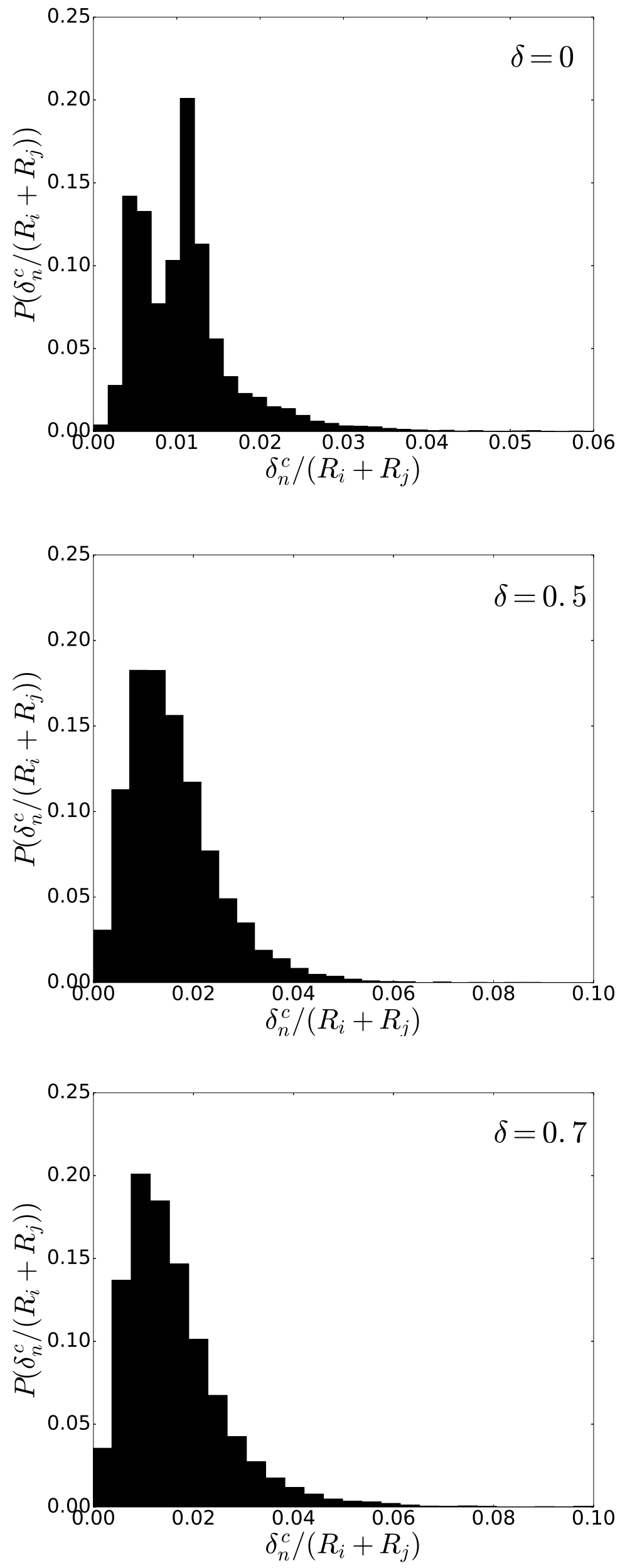

FIG. 14. Distribution of normal contact interpenetration for each packing of Fig.4b. Data are for loading state of the limit cycle (cycle 15). The friction between particles is $\mu = 0.3$.

## Appendix A: Distributions of contact interpenetration

Our simulations create initially loose packings with mean coordination number around $Z \sim 4$ for monodisperse system and slightly below this value for our most polydisperse case. The former result is consistent with the rule $Z_c = 2D$ applicable to frictionless monodisperse grains. Compressing the packings up to the maximum deformation imposed will increase the mean coordination number above $Z_c$.

We quantify the distribution of normal contact interpenetration for each particular packing presented in Fig.4b, corresponding to the limit cycle (compression state). This is achieved to explore the overlap at contacts between grains since it is known that in standard simulations of granular systems a maximum overlap of 1% is required in order to suppose a realistic model of the nature. We define the interpenetration at each contact as $\delta_n^c/(R_i + R_j)$, where $\delta_n^c$ is the normal overlaping and $R_i$ is the radii of each grain forming the contact. Figure 14 shows that the bulk of the grains for all values of polydispersity are below of 2% overlaping although few grains of the sample exhibit larger values. We think that those large values of interpenatration are responsible for the large and low values of the mean coordination number and porosity of the packings.

## Appendix B: Effect of particle angularity on macroscopic friction

In this appendix we briefly describe a simple model to contemplate angularity into our simulations and see how the trend between local and global friction is changed compared to that of rotating disks. We will show that within this model, "angular grains" approach the behaviour of non-rotating disks when angularity degree is increased. Particle angularity was introduced by clumping three disks together to construct a single angular grain, where its angularity degree can be characterized by changing the distance between particles.

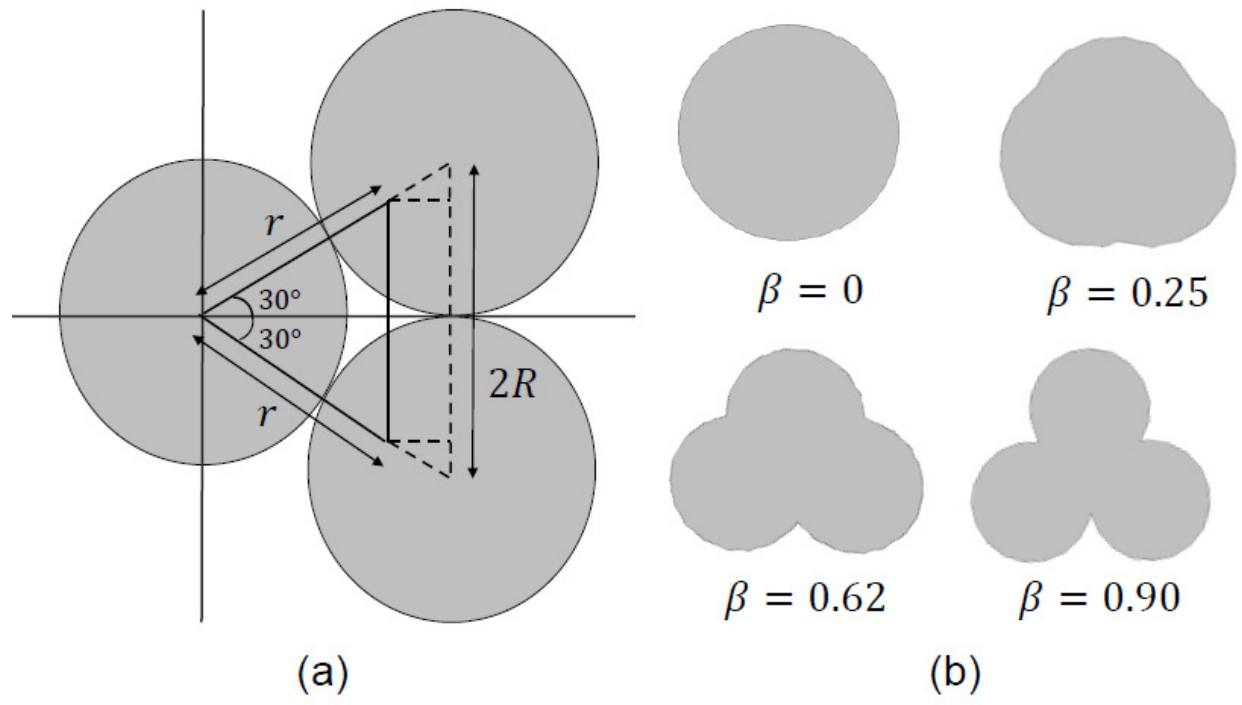

FIG. 15. (a) Model to create a grain with a particular angularity. (b) Four grains types with different angularity, characterized by the distance between particles $r$ in a triangular arrangement. As $\beta$ increases the particle angularity is higher.

To create a grain with a particular angularity, we start by fixing the center of one disk at the origin, then two more disks are added touching the first one with their centers at 30° angles with respect to the origin as indicated in Fig.15(a). This configuration represents the starting point of our particle angularity model, here $r = 2R$ and the angularity is a maximum. Now one can shorten $r$ with respect to $R$ and reduce $r$ all the way to zero, keeping the relative angles, reaching the limit of a circular grain. The angularity, which we label $\beta$ can then be quantified by the expression $\beta = \frac{r}{2R}$, for $0 < r < 2R$. Fig.15(b) shows different particle angularities constructed by our model. A similar approach for constructing angular grains was presented in [49, 50].

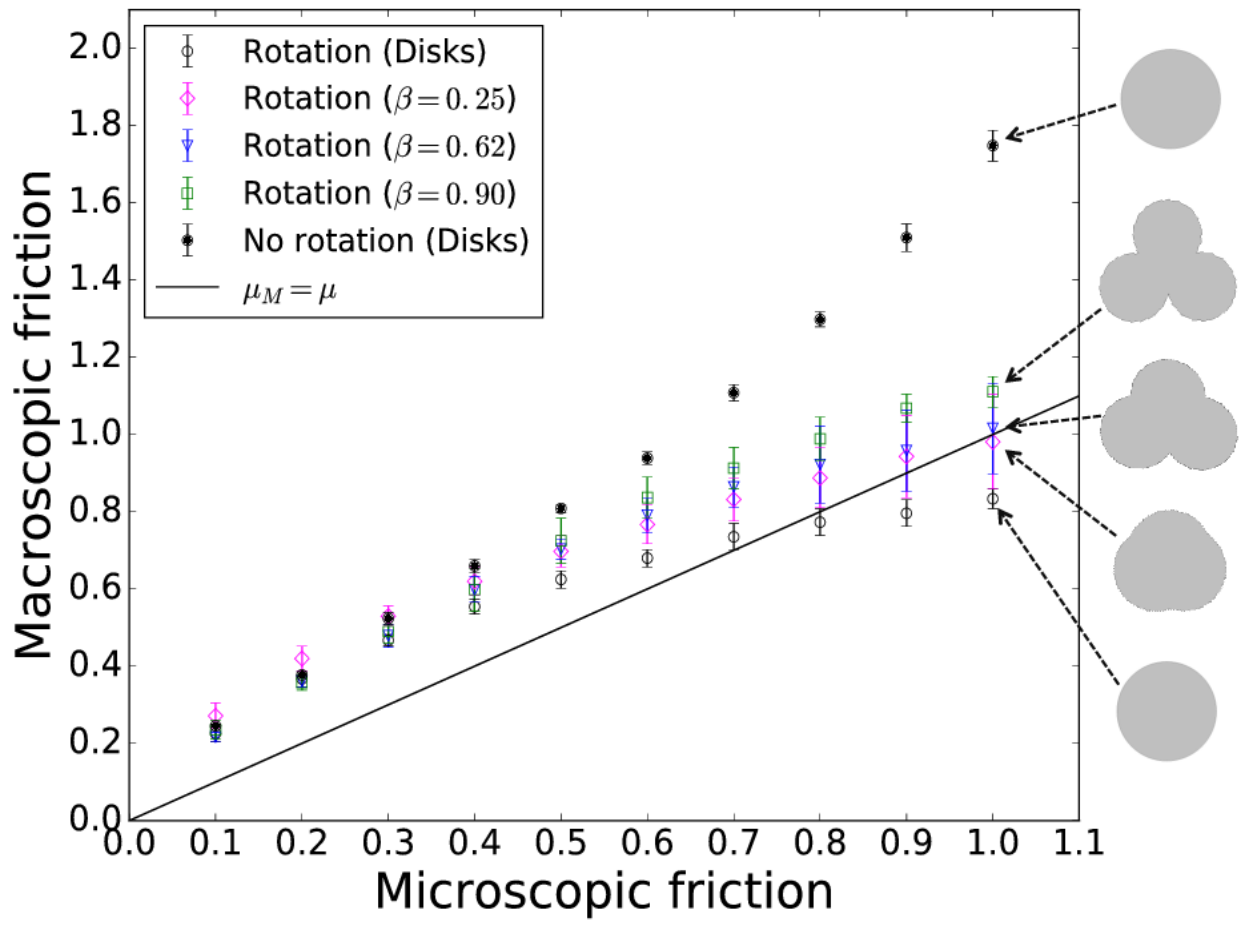

FIG. 16. Macroscopic friction as a function of particle friction when particle angularity changes. Data are shown for disks: Non rotating disks (filled symbol) and rotating disks (hollow symbols). We consider particle angularities with $\beta = 0.25$, $\beta = 0.62$ and $\beta = 0.90$.

The three composite particles with different angularities are used to construct three packings with a polydispersity of $\delta = 0.5$ in order to compare with those simulations obtained for disks. The composite particles contemplate rotations. The macroscopic friction for each packing was obtained as a function of the particle friction value. Figure 16 shows the results for the macro and micro friction when particle angularity changes. For low local friction, we obtain that all data fall on the curve corresponding to *non-rotating disks*, showing that particle sliding dominates over rotations. However, when particle friction increases, particle rotations become important giving rise to the saturation range exhibited by the macroscopic friction (as for rotating disks). Nevertheless, the macroscopic friction increases for large values of particle friction, as angularity increases. We can see that the range of agreement with the non-rotating disk curve increases. This suggests that frustration of particle rotations increases as angularity increases. These results are similar to previous works [49, 50], where they found that macro friction and mean friction mobilization increases when angularity increases. Such results justify the consideration of non-rotating disks in our simulations as a limiting model for angular grains. Even though rotating non-spherical particles would represent a more realistic representation of granular matter, we think that mapping the rolling resistance of disks as particle angularity [53] or preventing rotation, as in the present study (see also [13, 43]), will allow for representing well experimental results and remains much less computational effort than considering the rotational degree of freedom.